\documentclass[11pt, letterpaper]{article}

\usepackage[latin1]{inputenc} 
\usepackage[english]{babel}

\usepackage[dvips]{graphicx} 
\usepackage{amsmath}
\usepackage{graphics}
\usepackage{epsfig}
\usepackage{latexsym}
\usepackage{amssymb}
\usepackage{color}
\usepackage{url}
\usepackage{fullpage}

 \newtheorem{theorem}{Theorem}[section]

 {\refstepcounter{figure}}
 {\smallskip}

 { \begin{list}%
         {--}%
         {\setlength{\labelwidth}{20pt}%
          \setlength{\leftmargin}{25pt}%
          \setlength{\topsep}{0pt}
          \setlength{\itemsep}{0pt}
          \setlength{\parsep}{0pt}}}%
 { \end{list} }

 \newenvironment{remark*}
         {\bigskip\noindent \textbf{Remark.} \hspace{0.3mm}} 
         {\medskip} 

 \newenvironment{remarks*}
         {\bigskip\noindent \textbf{Remarks.} \hspace{0.3mm}} 
         {\medskip} 

 \newcommand{\qed}{$\square$}

 \newenvironment{proof}
         {\smallskip\noindent \textbf{Proof.} \hspace{0.3mm}} 
         {\qed  \medskip}

 \newcommand{\e}{\varepsilon}

 \newcommand{\sss}{{M}}
 \newcommand{\sssp}{\sss^+}
 \newcommand{\sssm}{\sss^-}
 \newcommand{\delsp}{\delsarg{\sssp}}
 \newcommand{\delsm}{\delsarg{\sssm}}

 \newcommand{\dual}[1]{\textrm{V}(#1)}

 \newcommand{\aff}{{\rm aff}}

 \newcommand{\R}{\mathbb{R}}

 \newcommand{\rch}{{\rm rch}}

 \newcommand{\del}{{\cal D}}

 \newcommand{\delsarg}[1]{\del^{#1}} 
 \newcommand{\dels}{\delsarg{\sss}}

 \newcommand{\tab }{\hspace{0.25cm} \begin{minipage}{7cm}} 
 \newcommand{\fintab }{\end{minipage} \par} 
 \newcommand{\algo}{\begin{center}\parbox{16cm}} 
 \newcommand{\finalgo}{\end{center}}

 \newcommand{\simplex}{\sigma}

 \newcommand{\complex}{{\cal C}}

 \newcommand{\wit}{W}
 \newcommand{\eee}{{L}}
 \newcommand{\winf}[1]{\complex^{#1}}
 \newcommand{\winfw}{\winf{\wit}}

\title{On the Topology of the Restricted Delaunay Triangulation 
and Witness Complex in Higher Dimensions} 
\author{Steve Y. Oudot}
\date{Submitted 30th November 2006, Revised 9th March 2008}

\begin{document}

\maketitle

\begin{abstract}
It is a well-known fact that, under mild sampling conditions, the
restricted Delaunay triangulation provides good topological
approximations of 1- and 2-manifolds. We show that this is not the
case for higher-dimensional manifolds, even under stronger sampling
conditions. Specifically, it is not true that, for any compact closed
submanifold $\sss$ of $\R^n$, and any sufficiently dense uniform
sampling $\eee$ of $\sss$, the Delaunay triangulation of $\eee$
restricted to $\sss$ is homeomorphic to $\sss$, or even homotopy
equivalent to it. Besides, it is not true either that, for any
sufficiently dense set $\wit$ of witnesses, the witness complex of
$\eee$ relative to $\wit$ contains or is contained in the restricted
Delaunay triangulation of $\eee$.
\end{abstract}

\section{Background and definitions}
\label{sec-defs}

All manifolds considered in this paper are compact closed submanifolds
of Euclidean space $\R^n$. The reach of a manifold $\sss$, or
$\rch(\sss)$ for short, is the minimum distance of a point on $\sss$
to the medial axis of $\sss$. All our manifolds have a positive
reach. This is equivalent to saying that they are $C^1$-continuous,
and that their normal vector field satisfies a Lipschitz
condition. 

Given a (finite or infinite) subset $\eee$ of a manifold $\sss$, and a
positive parameter $\e$, $\eee$ is an $\e$-sample of $\sss$ if every point
of $\sss$ is at Euclidean distance at most $\e$ to $\eee$. In
addition, $\eee$ is $\e$-sparse if the pairwise Euclidean distances
between the points of $\eee$ are at least $\e$. Note that an
$\e$-sparse sample of a compact set is always finite. Parameter $\e$
is sometimes made adaptative in the literature \cite{ab-srvf-99}, its
value depending on the distance to the medial axis of the manifold. In
this context, our $\e$-samples are called uniform $\e$-samples. 

For any finite set of points $\eee\subset\R^n$, $\del(\eee)$ denotes the
n-dimensional Delaunay triangulation of $\eee$, and $\dels(\eee)$ its
Delaunay triangulation restricted to a given subset $\sss$ of
$\R^n$. By definition, $\dels(\eee)$ is the nerve of the restriction
of the Voronoi diagram of $\eee$ to X. For any simplex $\simplex$ of
$\del(\eee)$, $\dual{\simplex}$ stands for the face of the Voronoi
diagram of $\eee$ that is dual to $\simplex$. The following result
comes from \cite{ab-srvf-99, abe-cbscc-98}: 
\begin{theorem}\label{th-curve-surf-homeo}
If $\sss$ is a smooth curve in the plane or a smooth surface in
3-space, and if $\eee$ is a finite $\e$-sample of $\sss$, with $\e <
0.1\;\rch(\sss)$, then $\dels(\eee)$ is homeomorphic to $\sss$.
\end{theorem}
Let $\eee$, $\wit$ be two subsets of $\R^n$, such that $\eee$ is
finite. Given a point $w\in\wit$ and a simplex $\simplex = [p_0,
  \cdots, p_k]$ with vertices in $\eee$, $w$ is a {\em witness} of
$\simplex$ (or simply $w$ {\em witnesses} $\simplex$) if $p_0,\cdots,
p_k$ are among the $k + 1$ nearest neighbors of $w$ in the Euclidean
metric, that is: $\forall p \in \{p_0,\cdots, p_k\}$, $\forall q \in
\eee \setminus \{p_0, \cdots, p_k\}$, $\|w-p\| \leq \|w-q\|$. The {\em
  witness complex} of $\eee$ relative to $\wit$, or $\winfw(\eee)$ for
short, is the maximum abstract simplicial complex with vertices in
$\eee$, whose faces are witnessed by points of $\wit$. From now on,
$\wit$ will be referred to as the set of witnesses, and $\eee$ as the
set of landmarks. As pointed out in \cite{ds-wdd-03, cds-teuwc-04},
when $\wit$ samples a manifold $\sss$, the witness complex
$\winfw(\eee)$ can be viewed as a discrete version of the restricted
Delaunay triangulation $\dels(\eee)$, and as such it should be closely
related to it. This is true indeed for curves and surfaces, as stated
in the following result of \cite{ae-wcrd-06, go-ruwc-07}:
\begin{theorem}\label{th-curve-surf-winf-dels}
There exists a positive constant $c$ such that, if $\sss$ is a smooth
curve in the plane or a smooth surface in 3-space, and if $\eee$ is an
$\e$-sample of $\sss$, with $\e\leq c\;\rch(\sss)$, then
$\winfw(\eee)$ is included in $\dels(\eee)$ for any set of witnesses
$\wit\subseteq\sss$, and it coincides with $\dels(\eee)$ if
$\wit=\sss$.
\end{theorem}

\section{Negative results}
\label{sec-neg-res}

\begin{figure}[tb]
\centering
\includegraphics[width=0.4\linewidth]{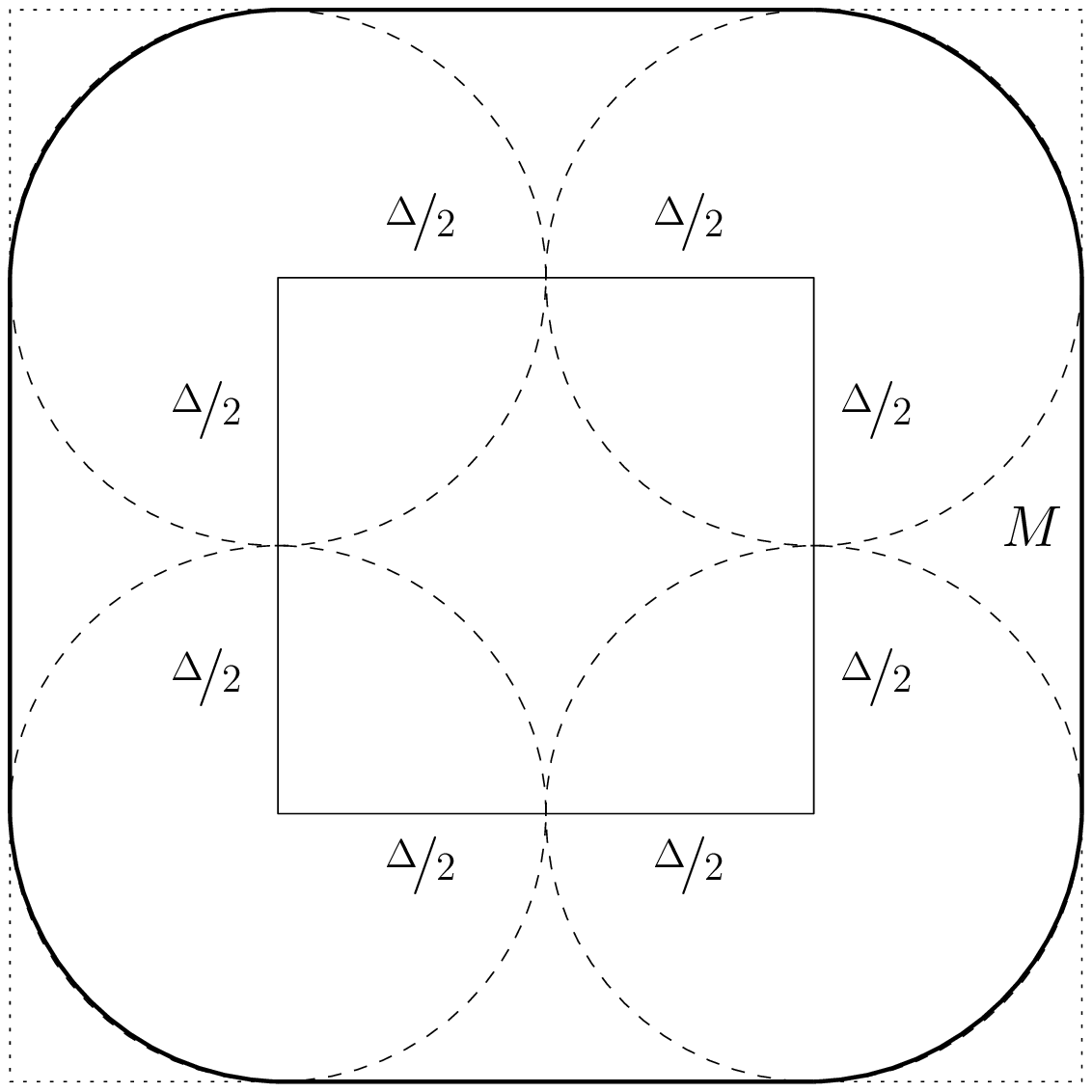}
\hspace{0.05\linewidth}
\includegraphics[width=0.4\linewidth]{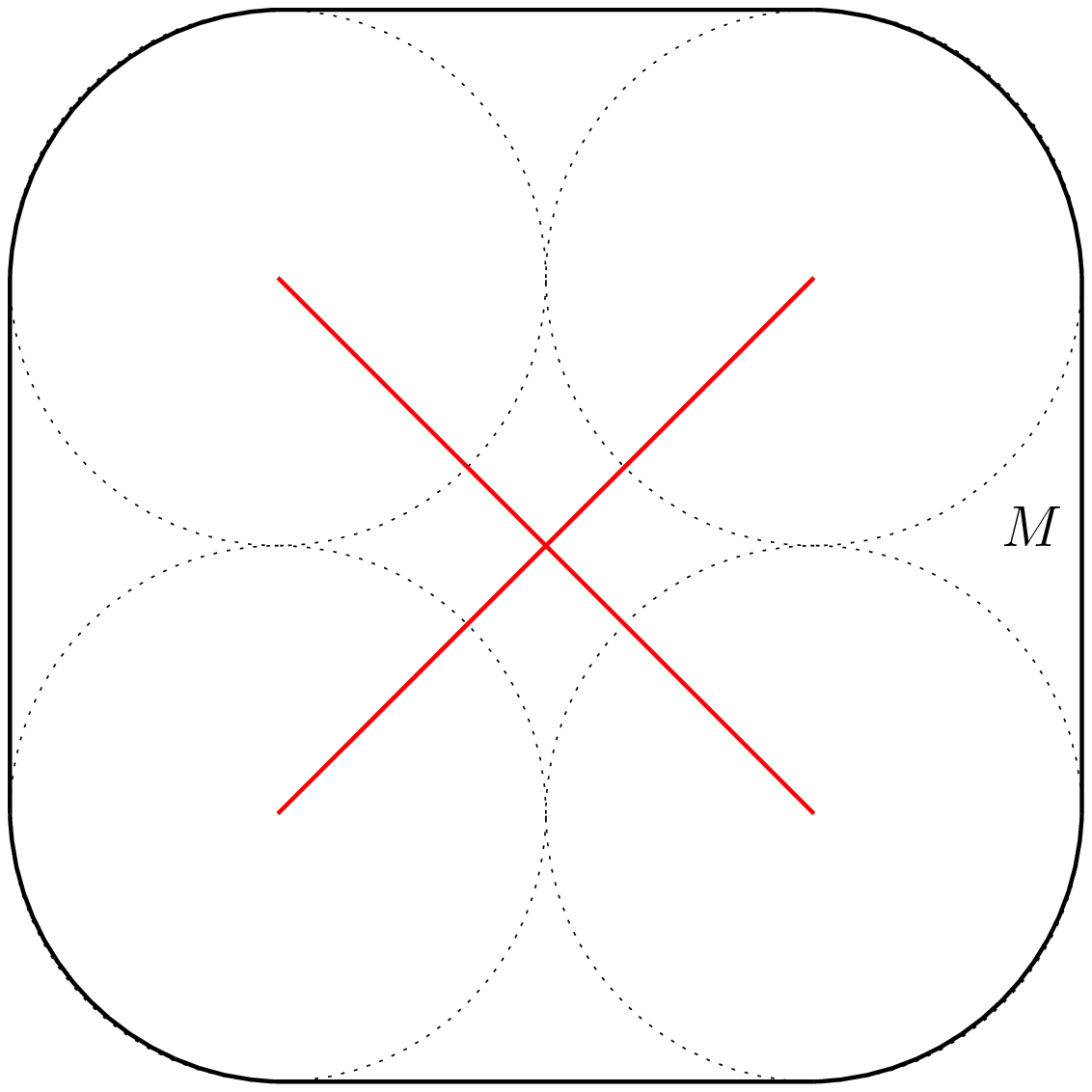}
\caption{Left: 2-d version of hypersurface $\sss$ (bold), defined as
  the boundary of the Minkowski sum of hypercube $[-\frac{\Delta}{2},
    \frac{\Delta}{2}]^2$ (solid) with the ball of radius
  $\frac{\Delta}{2}$ centered at the origin (copies of this ball are
  dashed). Hypercube $[-\Delta, \Delta]^2$ is marked by dotted
  lines. Right: $\sss$ and its medial axis.}
\label{fig:build-hypercube}
\end{figure}

In this section, we prove that Theorems \ref{th-curve-surf-homeo} and
\ref{th-curve-surf-winf-dels} do not hold as is for higher-dimensional
manifolds. We first show that $\dels(\eee)$ is not always homeomorphic
to $\sss$, even though $\eee$ is an $\Omega(\e)$-sparse $O(\e)$-sample
of $\sss$, for arbitrarily small $\e$ (Theorem
\ref{th-dels-not-homeo-S}). Our proof builds on an example of
\cite[\textsection11]{cdr-mrps-2005}, which deals with hypersurfaces
in $\R^4$. The intuitive idea is that, when $\dels(\eee)$ contains
badly-shaped tetrahedra, called {\em slivers}, it is possible to make
its normals turn by a large angle (say $\frac{\pi}{2}$) by perturbing
the points of $\eee$ infinitesimally. It follows that the
combinatorial structure of $\dels(\eee)$ can be modified by small
perturbations of $\sss$. We then extend our counter-example to show
that $\dels(\eee)$ may even not be homotopy equivalent to $\sss$
(Theorem \ref{th-dels-not-homeq-S}). Finally, we show that
$\winfw(\eee)$ may not be included in $\dels(\eee)$, even for
arbitrarily dense sets $\wit\subseteq\sss$ (Theorem
\ref{th-winf-notin-dels}). The fact that $\winfw(\eee)$ may not
contain $\dels(\eee)$ if $\wit\subsetneq\sss$ has already been proved
in \cite{go-ruwc-07}.
\begin{theorem}\label{th-dels-not-homeo-S}
For any positive constant $\mu < \frac{1}{3}$, there exist a compact
closed hypersurface $\sss$ in $\R^4$ and an $\Omega(\e)$-sparse
$O(\e)$-sample $\eee$ of $\sss$, with $\e = \mu\;\rch(\sss)$, such
that $\dels(\eee)$ is not homeomorphic to $\sss$. The constants hidden
in the $\Omega$ and $O$ notations are absolute and do not depend on
$\mu$.
\end{theorem}
\begin{proof}
Let $\Delta = \frac{2}{\mu}$. In $\R^4$, endowed with an orthonormal
frame $(x, y, z, t)$, we construct a hypersurface $\sss$ of reach
$\frac{\Delta}{2} =\frac{1}{\mu}$. Consider the Minkowski sum of
hypercube $[-\frac{\Delta}{2}, \frac{\Delta}{2}]^4$ with the ball of
radius $\frac{\Delta}{2}$ centered at the origin. The result is a
smoothed-out version of hypercube $[-\Delta, \Delta]^4$, as
illustrated in Figure \ref{fig:build-hypercube} (left).  Let $\sss$ be
its boundary. The reach of $\sss$ is $\frac{\Delta}{2}$, as shown in
Figure \ref{fig:build-hypercube} (right). Let $\e = \mu\;\rch(\sss) =
1$, and let $\delta > 0$ be an arbitrarily small parameter.
Consider points $u = (1, 0, 0, \Delta)$, $v = (1, 1, 0, \Delta)$, $w =
(0, 1, 0, \Delta)$, and $p_0 = (0, 0, \delta, \Delta)$. Let $c_0 =
(\frac{1}{2}, \frac{1}{2}$, $\frac{\delta}{2}, \Delta)$. It is easily
seen that $c_0$ is the circumcenter of $[u, v, w, p_0]$. Moreover, all
these points belong to $\sss$, which coincides with hyperplane $t =
\Delta$ in their vicinity. Let $r_0 = \|c_0 -u\| = \|c_0 -v\| = \|c_0
-w\| = \|c_0 -p_0\|$. We generate an $\e$-sparse $2\e$-sample $\eee_0$
of $\sss$ by an iterative process, starting with $\eee_0 = \{u, v, w,
p_0\}$, and inserting at each iteration the point of $\sss$ lying
furthest away from the current point set $\eee_0$, until the farthest
point of $\sss$ is no farther than $2\e$ from $\eee_0$. Since $\sss$
is compact, the process terminates, and the outcome is a $2\e$-sample
of $\sss$. Moreover, since $u, v, w, p_0$ lie at least $\e$ away from
one another, and since every point inserted in $\eee_0$ lies at least
$2\e$ away from $\eee_0$ at the time of its insertion, $\eee_0$ is
$\e$-sparse. Finally, no point of ball $B(c_0 , r_0 )$ lies farther
from $\{u, v, w, p_0 \}$ than $2r_0 = 2\sqrt{ \frac{1}{2}
  +\frac{\delta^2}{4}}$, which is less that $2\e$ since $\delta$ is
arbitrarily small. It follows that the interior of $B(c_0 , r_0 )$
contains no point of $\eee_0$, which implies that $[u, v, w, p_0 ]$
belongs to $\dels(\eee_0)$, its dual Voronoi edge intersecting $\sss$
at $c_0$. Observe also that, since $u, v, w, p_0$ belong to hyperplane
$t = \Delta$, the normal of $[u, v, w, p_0]$ is aligned with vector
$(0, 0, 0, 1)$, as shown in Figure \ref{fig:perturb-hypercube} (left).

\begin{figure}[tb]
\centering
\includegraphics[width=0.45\linewidth]{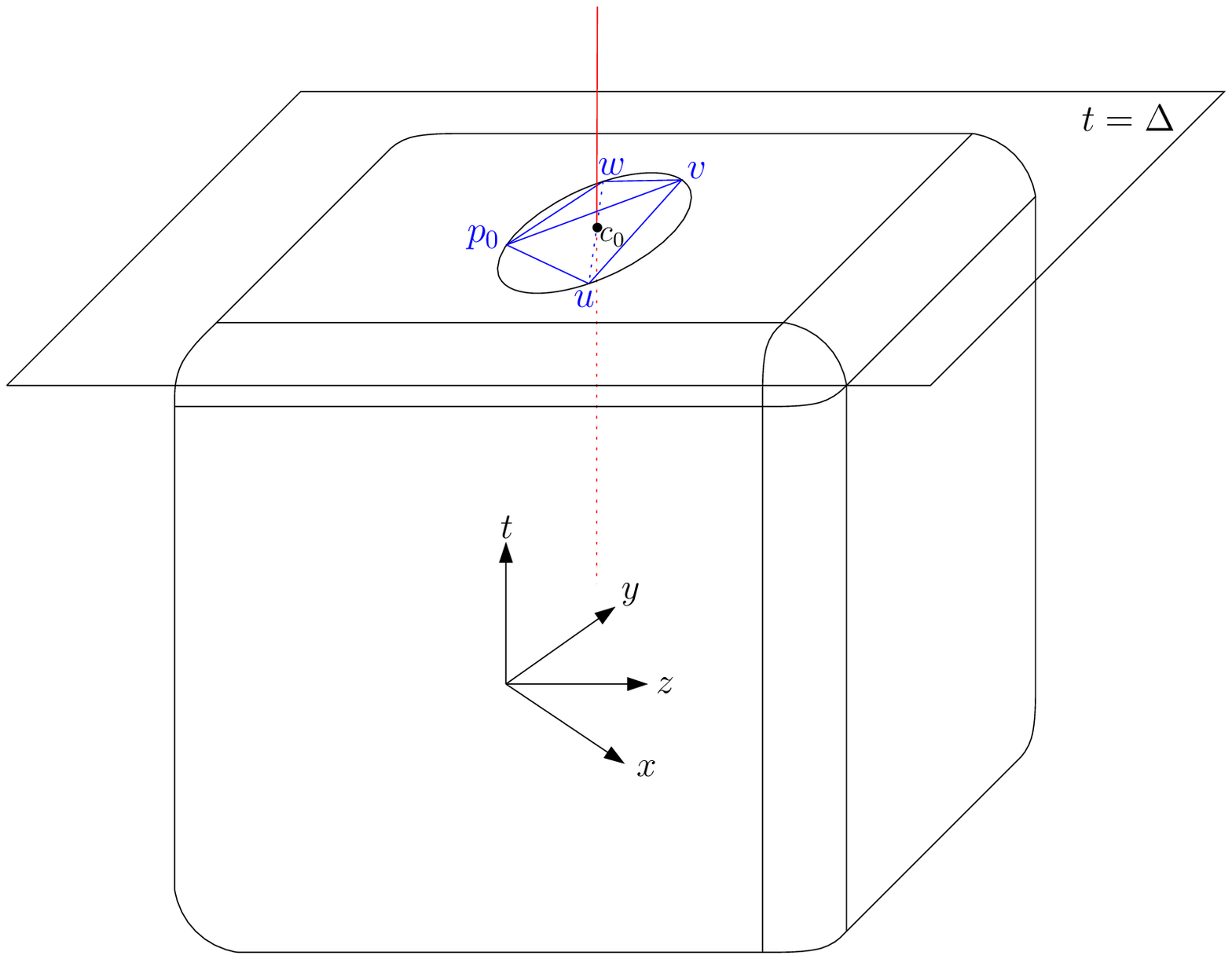}
\hspace{0.05\linewidth}
\includegraphics[width=0.45\linewidth]{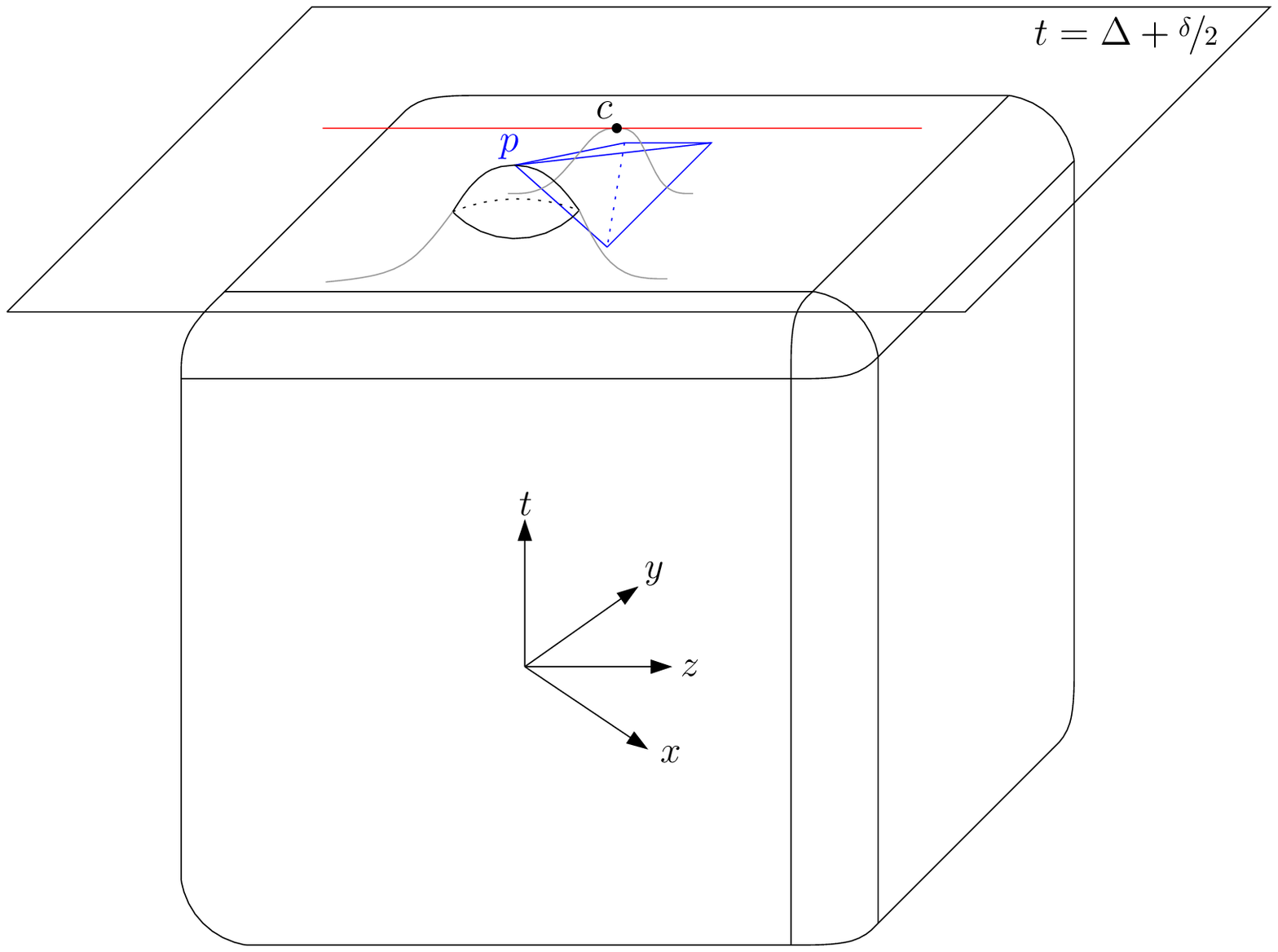}
\caption{Left: tetrahedron $[u, v, w, p_0]$ and its dual Voronoi
  edge. Right: after perturbation of $\sss$.}
\label{fig:perturb-hypercube}
\end{figure}

We now deform $\sss$ slightly and create a small bump at $c_0$, such
that the top of the bump is moved by $\frac{\delta}{2}$ into the
$t$-dimension, outward the hypercube. This bump changes the local
feature size of $\sss$. However, since $\delta$ is arbitrarily small,
the radius of curvature of the bump can be forced to be at least
$\frac{\Delta}{2}$, which implies that the reach of $\sss$ remains
equal to $\frac{\Delta}{2} = \frac{1}{\mu}$. Furthermore, since $c_0$
is the center of a Delaunay ball of radius greater than $1$, we can
assume that $\delta$ is small enough for the points of $\eee_0$ to
remain on $\sss$. Let $c = (\frac{1}{2}, \frac{1}{2},
\frac{\delta}{2}, \Delta+\frac{\delta}{2})$ be at the top of the
bump. Since the points of $\eee_0$ are located in hyperplane $t =
\Delta$ in the vicinity of $[u, v, w, p_0]$, $c$ is equidistant to $u,
v, w, p_0$, and closer to these points than to any other point of
$\eee_0$. This implies that the open ball $B_c = B(c,\|c-u\|)$
contains no point of $\eee_0$ and has $u, v, w, p_0$ on its bounding
sphere. Hence, $B_c$ is a Delaunay ball circumscribing $[u, v, w, p_0
]$, and $c$ belongs to the Voronoi edge dual to $[u, v, w, p_0
]$. Moreover, since $u, v, w$ and $(0, 0, 0, \Delta)$ are cocircular,
$\partial B_c$ passes also through $(0, 0, 0, \Delta)$.

We deform $\sss$ further by creating another small bump, at point $(0,
0, 0, \Delta)$ this time, so as to move this point by $\delta$ into
the $t$-dimension, outward the hypercube. Let $p = (0, 0, 0, \Delta +
\delta)$ be the top of the bump --- see Figure
\ref{fig:perturb-hypercube} (right). A quick computation shows that
$\|c-p\| = \|c-u\|$, which implies that $p\in\partial B_c $. Here
again, by choosing $\delta$ sufficiently small, we can make sure that
the radius of curvature of the bump is at least $\frac{\Delta}{2}$,
which means that the reach of the deformed hypersurface is still
$\frac{\Delta}{2} = \frac{1}{\mu}$. We can also make sure that the
bump of $p$ is disjoint from the bump of c since $\|c-p\| >
\frac{1}{\sqrt{2}}$, and that the points of $\eee_0\setminus\{p_0\}$
remain\footnote{They lie at least $\e$ away from $p_0$, and hence at
  least $\e-\delta$ away from $(0, 0, 0, \Delta)$.} on $\sss$. It
follows that $B_c$ is empty of points of $\eee$, where $\eee$ is
defined by $\eee = \eee_0 \cup \{p\} \setminus \{p_0\}$. Since
$\partial B_c$ contains $u, v, w, p$, $B_c$ is a Delaunay ball
circumscribing $[u, v, w, p]$. Equivalently, $c$ belongs to the
Voronoi edge $e$ dual to $[u, v, w, p]$. Note also that $\eee$ is an
($\e-\delta$)-sparse ($2\e+\delta$)-sample of $\sss$.

Since $[u, v, w, p]$ is included in hyperplane $z = 0$, its dual
Voronoi edge $e$ is aligned with $(0, 0, 1, 0)$, as illustrated in
Figure \ref{fig:perturb-hypercube} (right). This edge is incident to
four Voronoi 2-faces, which are dual to the four facets of $[u, v, w,
  p]$. These 2-faces can be seen as extrusions, into the $z$-dimension
$(0, 0, 1, 0)$, of the edges of the Voronoi diagram of $\{u, v, w,
p\}$ inside hyperplane $z = 0$. Among these Voronoi edges, two lie
above the plane $t = \Delta + \frac{\delta}{2}$, and two lie below. As
a result, in $\R^4$, two Voronoi 2-faces incident to $e$ lie above
hyperplane $t = \Delta + \frac{\delta}{2}$. These two Voronoi 2-faces
do not intersect $\sss$, except at $c$ and possibly at the bump of
$p$. Now, the circumradii of the facets of $[u, v, w, p]$ are at most
$\|c-u\| = \frac{\sqrt{1+\delta^2}}{\sqrt{2}} < \mu\;\rch(\sss)$,
thus, inside hyperplane $z = 0$, Amenta and Bern's normal lemma
\cite[Lemma 7]{ab-srvf-99} states that the edges of the Voronoi
diagram of $\{u, v, w, p\}$ make angles of at most
$\arcsin\frac{\mu\sqrt{3}}{1-\mu} < \frac{\pi}{3}$ with vector $(0, 0,
0, 1)$. As a consequence, any Voronoi 2-face $f$ incident to $e$ in
$\R^4$ makes an angle of at most $\frac{\pi}{3}$ with the plane
passing through $c$, of directions $(0, 0, 1, 0)$ and $(0, 0, 0, 1)$
(note that the affine hull $\aff(f)$ intersects this plane along the
line $\aff(e)$). Since $p$ lies $\frac{1}{\sqrt{2}}$ away from this
plane and only $\frac{\delta}{2}$ above $c$, for sufficiently small
$\delta$ the Voronoi 2-faces incident to $e$ lying above hyperplane $t
= \Delta + \frac{\delta}{2}$ do not intersect the bump of $p$. As a
consequence, they intersect $\sss$ only at $c$, and therefore their
dual Delaunay triangles are incident to exactly one tetrahedron of
$\dels(\eee)$, namely $[u, v, w, p]$. Hence, $\dels(\eee)$ is not a
closed hypersurface, and for this reason it cannot be homeomorphic to
$\sss$.
\end{proof}

Observe that the example given in the proof corresponds to a
degenerate case, since the Voronoi edge $e$ dual to tetrahedron $[u,
  v, w, p]$ intersects $\sss$ tangentially at $c$. This degeneracy can
be removed by inflating the bump of $c$ infinitesimally, so that it
intersects $e$ twice and transversally, but still no other Voronoi
edge.

Note also that tetrahedron $[u, v, w, p]$ is a
sliver, since vertex $p$ lies close to the affine hull of $[u, v,
  w]$. The original counter-example of \cite{cdr-mrps-2005} was
designed to highlight the fact that the normals of slivers in the
restricted Delaunay triangulation may differ significantly from the
normals of the underlying manifold. This is not true for non-sliver
simplices, as shown in Lemma 15 of \cite{cdr-mrps-2005}. Thus,
the fact that $[u, v, w, p]$ is a sliver in our
counter-example is crucial. 
\begin{theorem}\label{th-dels-not-homeq-S}
For any positive constant $\mu<\frac{1}{3}$, there exist a compact
closed hypersurface $\sss$ in $\R^4$ and an $\Omega(\e)$-sparse
$O(\e)$-sample $\eee$ of $\sss$, with $\e = \mu\;\rch(\sss)$, such
that $\dels(\eee)$ is not homotopy equivalent to $\sss$. The constants
hidden in the $\Omega$ and $O$ notations are absolute and do not
depend on $\mu$.
\end{theorem}
\begin{proof}
Let $\Delta = \frac{2}{\mu}$, and let $\delta > 0$ be an arbitrarily
small parameter. We begin our analysis with the example built in the
proof of Theorem \ref{th-dels-not-homeo-S}. We will modify $\sss$ and
$\eee$ in such a way that tetrahedron $[p, u, v, w]$ will no longer belong
to $\dels(\eee)$ while its four facets will still. This will prevent
$\dels(\eee)$ from being homotopy equivalent to $\sss$. 

\begin{figure}[tb]
\centering
\includegraphics[width=0.8\linewidth]{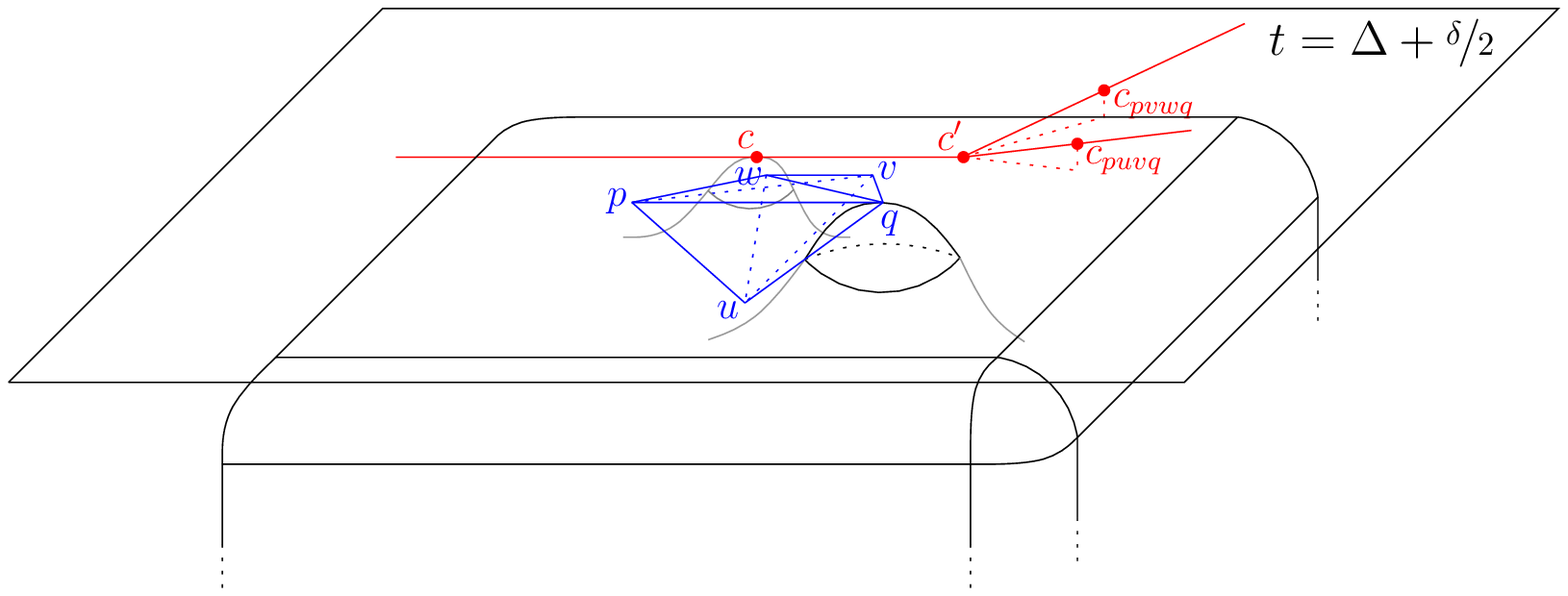}
\hspace{0.05\linewidth}
\includegraphics[width=0.8\linewidth]{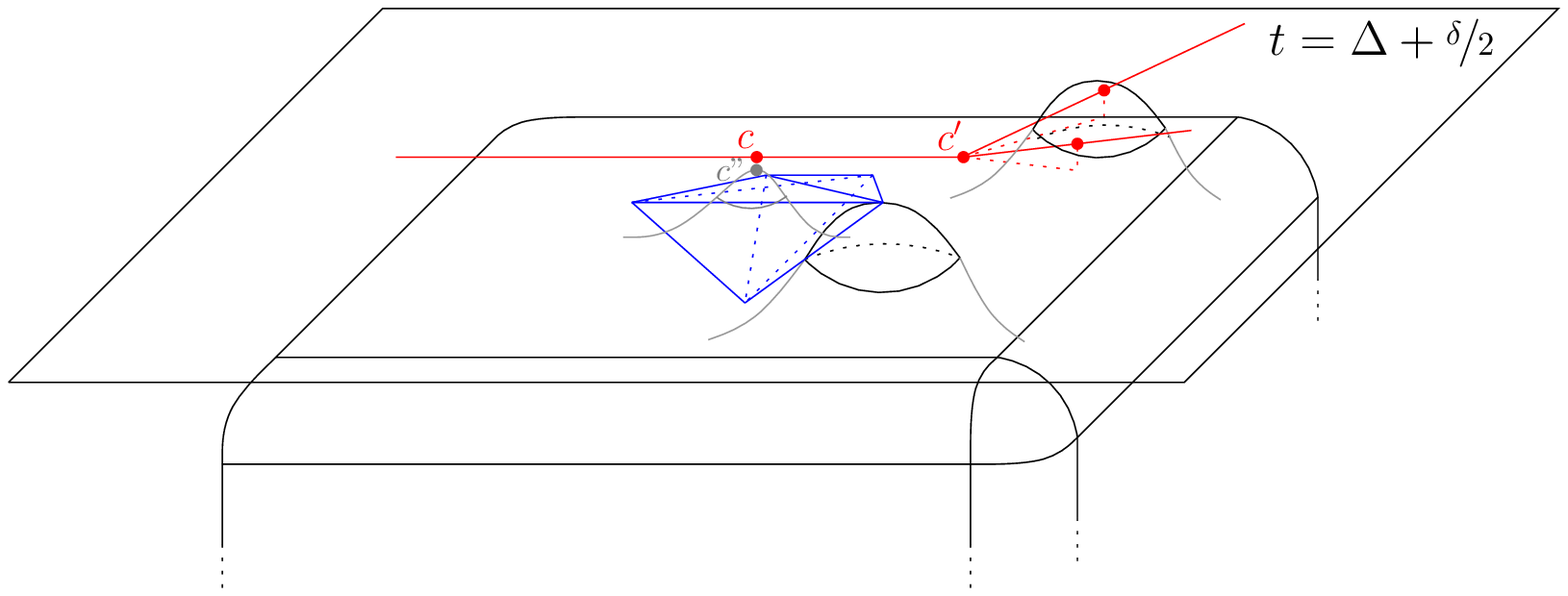}
\caption{Top: pentahedron $[p, u, v, w, q]$ and the duals of $[p, u, v,
    w]$, $[p, u, v, q]$, and $[p, v, w, q]$. Bottom: the bump of $c_{puvq}$ and
  $c_{pvwq}$ (the bump of $c$ has been slightly deflated).}
\label{fig:perturb-further-hypercube}
\end{figure}

Consider point $q = (\frac{1+\sqrt{2}}{2}, \frac{1}{2}, \delta^2,
\Delta+\delta)$. The distance of $q$ to hyperplane $t = \Delta$ is
$\delta$, which is arbitrarily small. Hence, as explained in the proof
of Theorem \ref{th-dels-not-homeo-S}, it is possible to deform $\sss$
slightly by creating a small bump of radius of curvature at least
$\frac{\Delta}{2}$ at point $(\frac{1+\sqrt{2}}{2}, \frac{1}{2},
\delta^2, \Delta)$, so that $\sss$ now passes through $q$ while its
reach remains $\frac{\Delta}{2}=\frac{1}{\mu}$. Moreover, since $q$
lies farther than $\frac{1}{2}$ from $\{p, u, v, w\}$, we can assume
without loss of generality that its bump does not affect the positions
of $p, u, v, w$.

The circumcenter of pentahedron $[p, u, v, w, q]$ is $c'=
(\frac{1}{2}, \frac{1}{2}, \frac{\delta^2}{2},
\Delta+\frac{\delta}{2})$, and its circumradius $r'$ is less than $\e
= 1$ (for sufficiently small $\delta$). It follows that every point of
$\sss$ lying in the ball $B(c', r')$ is at distance $O(\e)$ of $\{p, u,
v, w, q\}$. In addition, $q$ is farther than $\frac{\e}{2}$ from $\{p,
u, v, w\}$. Therefore, if we modify $\eee$ by inserting $q$ and
deleting all the points that lie in the interior of $B(c', r')$, $\eee$
remains an $\Omega(\e)$-sparse $O(\e)$-sample of $\sss$. Moreover,
$[p, u, v, w, q]$ is now a Delaunay pentahedron, whose dual Voronoi
vertex is $c'$.

Note that line $(c, c')$ is the affine hull of the Voronoi edge $e$
dual to $[p, u, v, w]$, and that $c'$ is an endpoint of $e$ --- see
Figure \ref{fig:perturb-further-hypercube} (top). Recall that, among
the four 2-faces incident to $e$, two lie above hyperplane $t = \Delta
+ \frac{\delta}{2}$. Let $f_{puv}$ and $f_{pvw}$ denote these two
2-faces. They are dual to triangles $[p, u, v]$ and $[p, v, w]$
respectively, since $p$ lies above hyperplane $t = \Delta$, which
contains $[u, v, w]$. Moreover, $f_{puv}$ and $f_{pvw}$ are convex
polygons, whose boundaries are two cycles of Voronoi edges that
intersect each other along $e$. In the cycle of $\partial f_{puv}$,
one edge adjacent to $e$, noted $e_{puvq}$, is dual to tetrahedron
$[p, u, v, q]$. Similarly, in the cycle of $\partial f_{pvw}$, one
edge adjacent to $e$, noted $e_{pvwq}$, is dual to $[p, v, w,
  q]$. Note that $c'$ is an endpoint of both $e_{puvq}$ and
$e_{pvwq}$. Moreover, it can be easily checked that the line
$\aff(e_{puvq})$ passes also through point $c_{puvq} =
\left(\frac{1}{2}+\frac{\delta^2(\delta^2+1)}{1+\sqrt{2}},
\frac{1}{2}, -\frac{1}{2},
\Delta+\frac{\delta}{2}+\frac{\delta(\delta^2+1)}{1+\sqrt{2}}\right)$,
while the line $\aff(e_{pvwq})$ passes through $c_{pvwq} =
\left(\frac{1}{2}, \frac{1}{2}+\delta^2(\delta^2+1), -\frac{1}{2},
\Delta+\frac{\delta}{2}+\delta(\delta^2+1)\right)$. This implies that
$e_{puvq}$ and $e_{pvwq}$ make angles of $O(\delta)$ with hyperplane
$t = \Delta+\frac{\delta}{2}$. So, we are in a situation where
tetrahedron $[p, u, v, w]$ has a horizontal dual edge, while two of
its adjacent tetrahedra, namely $[p, u, v, q]$ and $[p, v, w, q]$,
have almost horizontal dual edges, as illustrated at the top of Figure
\ref{fig:perturb-further-hypercube}.

Since $\|c_{puvq}-p\| = \|c_{puvq}-u\| = \|c_{puvq}-v\| =
\|c_{puvq}-q\| < \|c_{puvq}-w\|$, which is less than $\e = 1$ for
sufficiently small $\delta$, we can modify\footnote{For instance, we
  can simply delete the points of $\eee$ that lie in the interior of
  the ball $B(c_{puvq} , \|c_{puvq}-q\|)$.} $\eee$ such that the ball
$B(c_{puvq} , \|c_{puvq}-q\|)$ contains no point of $\eee$ in its
interior, while $\eee$ still remains an $\Omega(\e)$-sparse
$O(\e)$-sample of $\sss$. Similarly, we can assume without loss of
generality that $B(c_{pvwq} , \|c_{pvwq}-q\|)$ is a Delaunay ball. It
follows that $c_{puvq} \in e_{puvq}$ and $c_{pvwq} \in
e_{pvwq}$. Since $c_{puvq}$ and $c_{pvwq}$ lie $O(\delta)$ away from
each other, $O(\delta)$ above hyperplane $t = \Delta$, and
$\Omega(\e)$ away from $\eee$, we can deform $\sss$ by creating a bump
passing through $c_{puvq}$ and $c_{pvwq}$, of height $O(\delta)$ and
radius of curvature at least $\frac{\Delta}{2}$, while maintaining the
points of $\eee$ on $\sss$ --- see Figure
\ref{fig:perturb-further-hypercube} (bottom). Moreover, since
$c_{puvq}$ and $c_{pvwq}$ also lie $\Omega(\e)$ away from $c$, we can
assume without loss of generality that their bump does not touch the
bump of $c$. It follows that tetrahedra $[p, u, v, w]$, $[p, u, v,
  q]$, and $[p, v, w, q]$ belong to $\dels(\eee)$, while $\sss$ is
still tangent at $c$ to the Voronoi edge $e$ dual to tetrahedron $[p,
  u, v, w]$. We call $\sssp$ the current version of hypersurface
$\sss$, and $\delsp(\eee)$ the Delaunay triangulation of $\eee$
restricted to $\sssp$.

Our last operation consists in deflating slightly the bump of $c$,
such that $\sss$ no longer intersects $e$, and thus $[p, u, v, w]$ no
longer belongs to $\dels(\eee)$, as illustrated at the bottom of
Figure \ref{fig:perturb-further-hypercube}. Note however that $[p, u,
  v, q]$ and $[p, v, w, q]$ (and thus triangles $[p, u, v]$ and $[p,
  v, w]$) are still in $\dels(\eee)$, since the bump of $c_{puvq}$ and
$c_{pvwq}$ is disjoint from the bump of $c$. Recall also that the
Voronoi 2-faces dual to $[p, u, w]$ and $[u, v, w]$ lie below
hyperplane $t = \Delta + \frac{\delta}{2}$, and that they make angles
of at most $\frac{\pi}{3}$ with vector $(0, 0, 0, 1)$. Since the
deflation of the bump of $c$ is arbitrarily small, the Voronoi 2-faces
dual to $[p, u, w]$ and $[u, v, w]$ still intersect $\sss$. As a
consequence, the two triangles remain in $\dels(\eee)$. We call
$\sssm$ the current version of hypersurface $\sss$, and $\delsm(\eee)$
the Delaunay triangulation of $\eee$ restricted to $\sssm$.

The result of these operations is that, although $\eee$ is an
$\Omega(\e)$-sparse $O(\e)$-sample of both hypersurfaces $\sssp$ and
$\sssm$, whose homotopy types and reaches are the same as the ones of
$\sss$, $\delsp(\eee)$ and $\delsm(\eee)$ are different. Specifically,
tetrahedron $[p, u, v, w]$ is contained in $\delsp(\eee)$ but not in
$\delsm(\eee)$, whereas its facets belong to both complexes. It
follows that the Euler characteristics of $\delsp(\eee)$ and
$\delsm(\eee)$ differ\footnote{Specifically, $\chi(\delsp(\eee)) =
  \chi(\delsm(\eee))-1$.}, which implies that the complexes have
different homotopy types. Therefore, at least one of them is not
homotopy equivalent to the 3-sphere $\sss$.
\end{proof}

\paragraph{Witness complex.}
It is proved in \cite{go-ruwc-07} that the witness complex
$\winfw(\eee)$ may not contain $\dels(\eee)$ when the sets $\wit,\eee$
are drawn from a smooth surface $\sss$ such that
$\wit\subsetneq\sss$. However, we know from \cite{ae-wcrd-06} that
$\winfw(\eee)$ is still included in $\dels(\eee)$ in this case. Below
we prove that this latter statement no longer holds if $\sss$ is a
smooth manifold of dimension $3$ or more:
\begin{theorem}\label{th-winf-notin-dels}
 For any positive constants $\mu, \nu < \frac{1}{3}$, there exist a
 compact closed hypersurface $\sss$ in $\R^4$, an $\Omega(\e)$-sparse
 $O(\e)$-sample $\eee$ of $\sss$, and a $\delta$-sample $\wit$ of
 $\sss$, with $\e = \mu\;\rch(\sss)$ and $\delta = \nu\;\rch(\sss)$,
 such that $\winfw(\eee)$ is not included in $\dels(\eee)$. The
 constants hidden in the $\Omega$ and $O$ notations are absolute and
 do not depend on $\mu$ nor $\nu$. Moreover, $\wit$ can be made
 indifferently finite or infinite, and arbitrarily dense.
\end{theorem}
\begin{proof}
Let $\sssm$, $\eee$, $e$, and $c$ be defined as in the proof of
Theorem \ref{th-dels-not-homeq-S}. Recall that tetrahedron $[p, u, v,
  w]$ does not belong to $\delsm(\eee)$, whereas its facets do. We
assume without loss of generality that $c$ is not an endpoint of the
Voronoi edge $e$, which means that the bounding sphere of the Delaunay
ball $B(c, \|c-p\| )$ contains no point of $\eee$ other than $p, u, v,
w$. This condition can be ensured by an infinitesimal perturbation of
the points of $\eee\setminus \{p, u, v, w\}$. Let $d_c =
\min_{p'\in\eee\setminus\{p,u,v,w\}} \|c-p'\|$. This quantity is
greater than $\|c-p\|$ since $B(c, \|c-p\|)$ contains no point of
$\eee\setminus\{p, u, v, w\}$. 

Consider any (finite or infinite) set of witnesses
$\wit\subseteq\sssm$ such that, for each facet $\simplex$ of $[p, u,
  v, w]$, $\wit$ contains at least one point of
$\sssm\cap\dual{\simplex}$ (every such point witnesses $\simplex$ and
its subsimplices). Assume further that $\wit$ contains the top point
of the bump of $c$ (call this point $c''$). In the last stage of the
perturbation of $\sss$ described in the proof of Theorem
\ref{th-dels-not-homeq-S}, we slightly deflated the bump of $c$, such
that $c''$ lies strictly below $c$. Note that $p$ is the vertex of
$[p, u, v, w]$ lying furthest away from $c''$. Since the deflation is
arbitrarily small, we can assume without loss of generality that
$\|c-c''\|< \frac{1}{2} (d_c-\|c-p\|)$. This implies that the ball
$B(c'',\|c''-p\|)\subseteq B(c,\|c-p\|+2\|c-c''\|)$ is included in the
interior of $B(c, d_c)$. As a result, $B(c'',\|c''-p\|)$ contains no
point of $\eee\setminus\{p, u, v, w\}$. Since $p, u, v, w$ belong to
$B(c'',\|c''-p\|)$, tetrahedron $[p, u, v, w]$ is witnessed by
$c''$. And since the facets of $[p, u, v, w]$ and their subsimplices
are witnessed by points of $\wit$, $[p, u, v, w]$ belongs to the
witness complex $\winfw(\eee)$. However, we saw in the proof of
Theorem \ref{th-dels-not-homeq-S} that $[p, u, v, w]$ does not belong
to $\delsm(\eee)$.
\end{proof}

\section{Conclusion}
\label{sec-conclusion}

We have proved that the structural properties of the restricted
Delaunay triangualtion and witness complex on 1- and 2-manifolds do
not hold on higher-dimensional manifolds. This implies in particular
that the Delaunay-based approach to meshing and reconstruction is
unlikely to work as is in higher dimensions. One possible way of
getting rid of pathological cases is to use the sliver exudation
technique of \cite{cdeft-se-00}, which assigns weights to the vertices
of the triangulation in order to remove slivers from the vicinity of
the restrited Delaunay triangulation. This strategy has been
successfully applied in \cite{bgo-mradwc-07, cdr-mrps-2005}.

\bibliography{geom,geometrica} \bibliographystyle{plain}

\end{document}